\documentclass[twocolumn,english]{revtex4}
\usepackage[T1]{fontenc}
\usepackage[latin1]{inputenc}
\usepackage{float}
\usepackage{graphicx}
\usepackage{amsmath}
\usepackage{amsfonts}
\makeatletter

\usepackage{float}

\usepackage{graphicx}

\usepackage{babel}
\makeatother
\begin{document}

\title{Kondo temperature and crystal field levels of Ce systems within a finite U Non-Crossing Approximation}

\author{P. Roura-Bas~$^{a}$ ,V. Vildosola
$^{a}$, L.O. Manuel~$^{b}$~and A. M. Llois~$^{a,c}$}

\affiliation{$^{a}$ Dpto de F\'{\i}sica, Centro
At\'{o}mico Constituyentes,
Comisi\'{o}n Nacional de Energ\'{\i}a At\'{o}mica, Buenos
Aires, Argentina}

\affiliation{$^{b}$ Instituto de F\'{\i}sica Rosario,
Universidad Nacional de Rosario, Rosario, Argentina}

\affiliation{$^{c}$ Dpto de F\'{\i}sica,
Universidad de Buenos Aires, Buenos
Aires, Argentina}

\begin{abstract}

We calculate the Kondo temperature ($T_K$) and 
crystal-field levels of strongly correlated multiorbital systems 
solving the Anderson Impurity Model with the finite U Non-Crossing 
Approximation (UNCA) in its simplest scheme,  that is, considering
the self energies at lowest order in the $1/N$ diagrammatic 
expansion. We introduced an approximation to the vertex function that includes
the double energy dependence and investigate its effect on the values of $T_K$ for 
simple electronic  models. We also analyze the competition between 
the two spin flip mechanisms, involving virtual transitions to empty
and doubly occupied states, in  the 
determination of the ground state symmetry by including an extra 
diagram of higher order in $1/N.$
We finally combine the resulting simple formalism with {\it ab initio} 
calculated electronic structures to obtain $T_K$'s, 
ground states, and  crystal field splittings in excellent agreement 
with experimental results for two particular Ce compounds, namely 
CeIn$_3$ and CeSn$_3$.
\end{abstract}
\maketitle

\section{Introduction}

One of the most studied models for strongly correlated electron systems is the Anderson impurity model. This can  account for the Kondo regime, where conduction electrons scatter off a localized magnetic
impurity and form a local singlet.
Initially proposed as a model for local moment formation in metals, in recent years it has been extensively 
used to describe the physics of quantum dots and strongly correlated lattice systems,
mainly, heavy fermion compounds.

Among the several methods proposed to solve the Anderson impurity Hamiltonian, the so-called
non-crossing approximation (NCA),  in its lowest order self-consistent form occupies a special place due to its computational simplicity. NCA can be thought as a pertubative expansion with respect to $1/N$, where $N$ is the degeneracy of the impurity levels.
In particular, the NCA has been widely used in the last twenty years to solve
the Anderson Hamiltonian in the infinite $U$ limit, in which the double
occupancy of the impurity site is prohibited. This scheme was successfully applied to, both,
theoretical models \cite{NCA-bickers-cox,Bickers,Coleman} and
real materials \cite{Han,Svane1,paper-CeXY}, to analyse different physical properties,
like magnetic susceptibilities, crystal-field splittings and spectral properties.
Also, within this approximation, the out-of-equilibrium transport properties
of single and double quantum dot systems have been studied\cite{Aguado-Langreth,Wingreen}.

For $U\rightarrow\infty$ and large $N$ degeneracy of the local states,
NCA captures the Kondo energy scale and provides a qualitative description of the formation of the
Kondo resonance when the temperature approaches the Kondo regime
$(T\rightarrow T_K)$ from above \cite{Coleman}.
However, this is not the case for temperatures much lower than $T_K$ ($T \ll T_K$),
where NCA yields unphysical results. Furthermore, the $T_K$'s are often underestimated in the infinite $U$ limit.

The deficiencies that restrict the usefulness of NCA are greatly reduced if a finite on-site repulsion $U$ is considered together with the inclusion of vertex corrections \cite{Pruschke89}.
While NCA in the infinite $U$ limit contains all non-crossing diagrams up to the
$(1/N)^{(1)}$ order, with a finite $U$ there appear crossing diagrams
of order $(1/N)^{(0)}$ which have to be at least included
through vertex corrections. 

Recently, Sakai \textit{et al.} \cite{sakay} and J. Otsuki and Y. Kuramoto \cite{Otsuki} have
 shown that a finite-$U$ NCA (UNCA) that collects all the $(1/N)^0$ order diagrams yields the proper energy scale
of the Kondo effect. Within  their approximation, which also considers non-crossing diagrams of order $1/N$, the main
problem is how to achieve a complete solution of the integral equations for
the vertex corrections, that depend on two independent energy variables.

A more sofisticated extension of the NCA technique has been introduced by Haule \textit{et
al.}\cite{Haule}. This extension, known as the symmetrized finite-$U$ NCA
(SUNCA), treats the fluctuation processes into the empty and into the doubly
occupied intermediate states on an equal footing, by means of a proper symmetrization of the  vertex corrections.
Although SUNCA provides a correct energy scale, its practical computation is not easy, even for a simple structure of the conduction electron band.

In order to be applied to real materials, the scheme to solve the many-body Hamiltonian should be able to give
both, a good approximation to the  Kondo temperature as well as the correct ground state symmetry without implying an unreasonable computational effort under realistic conditions (multiorbital models, complex band electron structures, etc.).

In an earlier contribution, we have been able to understand and predict trends in the evolution of crystal-field
splittings by using the NCA in the infinite $U$ limit, with the hybridization function taken
from \emph{ab initio} electronic structure calculations \cite{paper-CeXY}.
The aim of this contribution is twofold, on the one side we introduce an approximation to vertex functions including the double energy dependence in two different ways within UNCA at lowest order in $1/N$ and study its influence on T$_K$. On the other hand, we analyze the importance of diagrams of different orders on the correct prediction of the ground state symmetry. Finally, we exemplify with two real Cerium compounds whose hybridization functions are obtained from first principles.

The paper is organized as follows. In section II, we introduce the Anderson
Hamiltonian and the auxiliary particle formalism. 
We also present the UNCA equations up to the $(1/N)^{(0)}$ order of
the diagrams. In section III we summarize the basic results for
the Kondo temperature implementing different treatments for the
vertex functions. In section IV, we investigate the crystal-field splittings
induced by hybridization and the symmetry of the ground state. Finally, in
section V, we apply the previous ideas on two real systems. The Kondo
temperature and the symmetry of the  ground state are obtained for CeIn$_3$
and CeSn$_3$. In section VI, we summarize and conclude. Some details of the calculations can be found 
in the Appendices.

\section{Auxiliary particle representation and the large-$N$ NCA}
The Anderson impurity model with finite $U$ is described by the Hamiltonian
\begin{displaymath}
\hat{H}=\sum_{k m} \epsilon_{km} ~\hat{c}_{km}^\dagger \hat{c}_{km}
\end{displaymath}
\begin{displaymath}
+\sum_m\epsilon_{m} \hat{f}_m^\dagger \hat{f}_m+U\sum_{m<n}\hat{N}_{m}\hat{N}_{n}
\end{displaymath}
\begin{equation}
+\sum_{km}\left(V_{km}  \hat{f}_m^\dagger \hat{c}_{km}+V_{km}^\ast 
\hat{c}_{km}^\dagger \hat{f}_m\right), 
\end{equation}
where the indices $m,n$ label the quantum numbers of the impurity levels. The operators $\hat{c}_{km}^\dagger,\hat{f}_m^\dagger$ create a conduction and a localized electron state, respectively, and $\hat{N}_{m}=\hat{f}_m^\dagger\hat{f}_m$ is the $f$-number operator. The last term represents the hybridization between conduction and localized electrons, and $V_{km}$ are the hybridization matrix elements. In the auxiliary particle approach to the Anderson impurity model, the local impurity states are represented by additional degrees of freedom, assumed to be created by pseudo-bosons and pseudo-fermions 
operators\cite{Coleman}. In order to take into account the effects of a finite value of the on-site Coulomb interaction, $U$, the local Hilbert subspace must contain the empty, single, and doubly occupied states, while states with higher occupancies can surely be neglected for intermediate to large values of $U.$ The local states are represented as follows: 
\begin{displaymath}
\left|0\right\rangle=\hat{b}^\dagger\left|vac\right\rangle,
\end{displaymath}
\begin{displaymath}
\left|m\right\rangle=\hat{s}_{m}^\dagger\left|vac\right\rangle,
\end{displaymath}
\begin{displaymath}
\left|mm'\right\rangle=\hat{d}_{mm'}^\dagger\left|vac\right\rangle,
\end{displaymath}
where $|vac>$ is the vacuum state for the auxiliary particle operators, $\hat{b}^\dagger$ is the ''light'' boson (empty state), $\hat{s}_{m}^\dagger$'s are the single 
pseudo-fermions (single-occupied states with energies $\varepsilon_m$), and 
$\hat{d}_{mm'}^\dagger$'s are the ''heavy'' bosons (which correspond to the doubly 
occupied states with energies $E_{mn}=\epsilon_{m}+\epsilon_{n}+U$). 
These heavy bosons satisfy 
the antisymmetric property $\hat{d}_{mm'}^\dagger=-\hat{d}_{m'm}^\dagger$.
The local physical electron operator, $\hat{f}_m$, can be written as
\begin{displaymath}
\hat{f}_m = \hat{b}^\dagger\hat{s}_{m}+\sum_{m'\neq m} \hat{s}_{m'}^\dagger\hat{d}_{mm'}.
\end{displaymath}
In this representation the Hamiltonian becomes
\begin{displaymath}
\hat{H}=\sum_{km} \epsilon_{km} ~\hat{c}_{km}^\dagger \hat{c}_{km}+ 
\sum_m\epsilon_{m}\hat{s}_{m}^\dagger\hat{s}_{m} 
\end{displaymath}
\begin{displaymath}
+\sum_{m<n}\left(\epsilon_{m}+
\epsilon_{n}+U\right)\hat{d}_{mn}^\dagger\hat{d}_{mn}
\end{displaymath}
\begin{displaymath}
+\sum_{km}\left(V_{km}\hat{s}_{m}^\dagger \hat{b}\;\hat{c}_{km}+ \rm{H.c.}\right)
\end{displaymath}
\begin{equation}
+\sum_{kmm'(m'\neq m)}\left(V_{km}\hat{d}_{mm'}^\dagger\hat{s}_{m'}\hat{c}_{km}
+ \rm{H.c.}\right).
\end{equation}
The auxiliary particle Hamiltonian is invariant under simultaneous, local $U(1)$ 
gauge transformations, $\hat{a}\rightarrow\hat{a} e^{i\phi(t)}$, where $\phi(t)$ is an arbitrary, time-dependent phase and $\hat{a}$ is any auxiliary particle operator. Due to this, there is a conservation of local charge $Q$ in time.
Therefore, all physical quantities should be obtained with the constraint $Q=1$ \cite{Coleman}.
The corresponding Green's functions of the auxiliary particles have the usual structure $\mathcal{G}_{a}^{-1}(z)=z-\epsilon_{a}-\Sigma_{a}(z),$ and the self-energies $\Sigma_{a}(z)$ can be evaluated by means of the non-crossing approximation.  

In the context of the NCA, the simplest picture that contains the two elementary ''spin-flip'' scattering processes, involving empty and doubly occupied intermediate 
states, respectively, and that captures the correct energy scale of the Kondo effect in the
finite $U$ case,
is obtained with the inclusion of vertex corrections in a large-$N$ expansion, retaining only the lowest order diagrams in $(1/N)$, that is, $(1/N)^{(0)}.$\cite{Kang-Min}.  
In this approximation, called from now on UNCA$^{(0)}$, the self-energy of the heavy boson propagator vanishes. Consequently, the Green's function of the doubly occupied states can be written in the 
simple form  $\mathcal{G}_{mm'}^{-1}(z)=z-E_{mm'}.$ 
On the other hand, the self-energy for the pseudo-fermion propagators contain only  the contribution coming from the heavy boson. The crucial advantage of this scheme is that the system of equations does not have to be solved in a self-consistent way.
\begin{figure}[!h]
\begin{center}
\includegraphics[scale=0.5,angle=0]{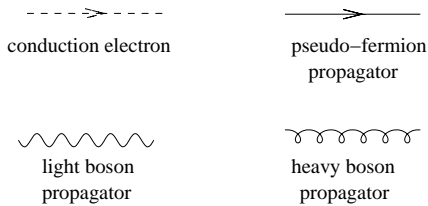}
\caption{Throughout this article, the full, wiggly, dashed, and curly lines 
stand for pseudofermion, empty boson, conduction electron, and heavy boson propagators, respectively. \label{nomenclatura}}  
\end{center}
\end{figure}
\begin{figure}[!h]
\begin{center}
\includegraphics[clip,scale=0.6,angle=0]{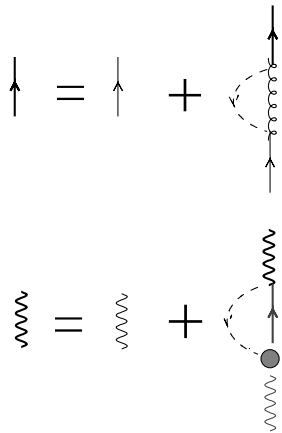}
\caption{Dyson's equation for the pseudo-fermion and light boson propagators within the non-crossing 
approximation up to $(1/N)^0$ order.
The lines used in Feynman diagrams are shown in Fig. \ref{nomenclatura} 
The bold (light) lines represent the full (free) propagators. 
The big dot represents the vertex function, that is, the bare hybridization vertex $V_{km}$ dressed by vertex corrections.  
\label{cap:unca0}}
\end{center}
\end{figure}
\begin{figure}[!h]
\begin{center}
\includegraphics[clip,scale=0.25,angle=0]{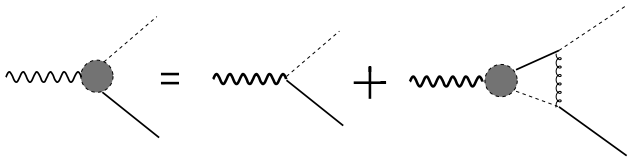}\end{center}
\caption{Diagrammatic representation of the integral equation for the vertex function in 
the ladder approximation. The vertex corrections incorporate crossing diagrams of order $(1/N)^0$ to the NCA equations.\label{vertex}}
\end{figure}
\\
The self-energies, denoted by loops in the diagrams of Fig. \ref{cap:unca0}, 
can be evaluated by standard Feynman rules
and by the projection process into the $Q=1$ subspace\cite{Costi_Kroha_Wolfle}. The vertex function $\Lambda_{m}(\omega,\epsilon)$, in the ladder approximation shown in Fig. \ref{vertex}, can also be evaluated in the same way. 
The final result is the following set of non-coupled equations:\\
\begin{equation} 
\label{eq:self-energy-fermion}
\Sigma_{m}{(\omega)=\sum_{m'\neq m}\int\frac{d\epsilon}{\pi}
~n_{F}(\epsilon)\Gamma_{m'}(\epsilon)\mathcal{G}_{mm'}(\omega+\epsilon)},
\end{equation}

\begin{equation} \label{eq:vertex-functions}\begin{split}
\Lambda_{m}(\omega,\epsilon)=1+&\sum_{m'\neq m} \int\frac{d\epsilon'}{\pi}
\Lambda_{m'}(\omega,\epsilon')n_{F}(\epsilon')\Gamma_{m'}(\epsilon')\times\\
&\times\mathcal{G}_{m'}(\omega+\epsilon')\mathcal{G}_{mm'}(\omega+\epsilon'+\epsilon),
\end{split}
\end{equation}
\begin{equation}
\label{eq:self-energy-boson}
\Sigma_{b}(\omega)=\sum_{m}\int\frac{d\epsilon}{\pi}~n_{F}
(\epsilon)~\Gamma_{m}(\epsilon)\Lambda_{m}(\omega,\epsilon)\mathcal{G}_{m}
(\omega+\epsilon).
\end{equation}
Here, $n_{F}(\epsilon)$ is the Fermi function and $\Gamma_{m}(\epsilon)$ are the hybridization functions between the conduction electron band and the impurity state, \begin{equation}
\label{eq:hybridization-functions}
\Gamma_{m}(\epsilon)=\pi\sum_{k}V_{km}V_{km}^\ast\delta(\epsilon-\epsilon_{km}). 
\end{equation}
It is easy to see that, with the self-energies for the pseudo-fermions computed from 
Eq. (\ref{eq:self-energy-fermion}), 
one can obtain the vertex functions through Eq. (\ref{eq:vertex-functions}). Finally, the self-energy for the light boson propagator is obtained directly 
from Eq. (\ref{eq:self-energy-boson}). 
A complete solution of the equations for the vertex functions seems to be impractical. In the next section we introduce some simplifications in order to solve them numerically with 
low computational effort, even for realistic systems.

\section{Approximations to the vertex function}

The vertex functions depend on two independent energy variables, $\omega$ and $\epsilon$. A complete solution of the set of integral equations 
(\ref{eq:vertex-functions}) implies a lot of computational effort, hence some simplification is needed. The schemes   NCA and UNCA$^{(0)}$ are conserving, that is, the diagrams can be derived from a Kadanoff-Baym functional\cite{kadanoff-baym,baym}. However, this property  is lost when  approximating  the vertex function. This non-conserving feature, does not imply major consequences if the spectral functions of the slave particles are always positive and also integrate to one, as expected for the conservation of the expectation value of the number of auxiliary particles in this theory ($Q$). We check that this is the case in the following calculations.   

In this section we compare the simplest approximations to solve numerically the vertex equations and analyse their influence on the calculated Kondo temperature for different values of the on-site Coulomb repulsion $U$.\\

\subsection{NCA$f^2$v approximation}

Sakai \textit{et al}\cite{sakay} have pointed out that the exchange coupling 
due to virtual transitions to the doubly occupied state can be obtained even when 
the $\epsilon$ energy dependence of the vertex function $\Lambda_m(\omega,\epsilon)$ is neglected. So, in order to get rid of the variable $\epsilon$, they replace the heavy boson propagator in Eq. (\ref{eq:vertex-functions}) by a representative value, that is, $\mathcal{G}_{mm'}(z\sim \epsilon_{m'})=-1/(\epsilon_{m}+U)$. Within this approximation there is no more $\epsilon$ dependence in the vertex functions:
\begin{equation}
\Lambda_{m}(\omega,\epsilon)\rightarrow\Lambda_{m}^{(a)}(\omega).
\end{equation}
In this way, Eq. (\ref{eq:vertex-functions}) is simplified as follows
\begin{equation}
\Lambda_{m}^{(a)}(\omega)+\sum_{m'\neq m}\frac{\Lambda_{m'}^{(a)}(\omega)}{\epsilon_m+U}F_{m'}(\omega)=1,
\end{equation}
where 
\begin{equation} \label{eq:F_{m}}
F_{m}(\omega)=\int\frac{d\epsilon}{\pi}~n_{F}(\epsilon)\Gamma_{m}(\epsilon)
\mathcal{G}_{m}(\omega+\epsilon). 
\end{equation}
With this simplification of the vertex functions, $\Lambda_{m}^{(a)}$, the boson self-energy 
turns into the simple expression
\begin{equation}
\Sigma_{b}^{(a)}(\omega)=\sum_{m}\Lambda_{m}^{(a)}(\omega)F_{m}(\omega).
\end{equation}
\\
Sakai \textit{et al.} have called their approximation NCA$f^2$v and, using it, 
they have obtained good results for different properties of real systems, like CeSb\cite{sakay}. On the other hand, Svane \textit{et al.}\cite{Svane}
have also used vertex functions with a single energy argument to analyse the monopnictides series of Cerium compounds, CeX (X=N, P, As, Sb, Bi).\\

\subsection{NCA$f^2$v$\epsilon$ approximation}

We can recover the second energy dependence, $\epsilon$, in the vertex functions, 
avoiding at the same time to solve the integral equations, if we 
interchange energy variables in the integrand of Eq. (\ref{eq:vertex-functions}) 
as follows: 
\begin{equation} 
\label{eq:vertex(b)}
\Lambda_{m'}(\omega,\epsilon')\rightarrow\Lambda_{m'}^{(b)}(\omega,\epsilon) .
\end{equation}
This procedure is justified under the assumption that the $\epsilon$ dependence
of the vertex functions is relatively weak. 
The resulting approximation for the vertex functions was successfully used by Kang \textit{et al.}\cite{Kang-Min} to study the influence of a magnetic field in the 
impurity Anderson model. Even though we recover the double energy dependence of the 
vertex functions, it should be noticed that energy conservation is still not completely 
satisfied in the vertex correction diagrams. Under this approximation, the vertex
equations (\ref{eq:vertex-functions}) become a set of linear algebraic equations for 
given values of $\omega$ and $\epsilon$, 
\begin{equation}
\label{eq:ncaf2ve}
\begin{split}
\Lambda_{m}^{(b)}(\omega,\epsilon)=&1+\sum_{m'\neq m}\Lambda_{m'}^{(b)}(\omega,\epsilon)\int\frac{d\epsilon'}{\pi}~n_{F}(\epsilon')\times\\
&\Gamma_{m'}(\epsilon')\mathcal{G}_{m'}(\omega+\epsilon')\mathcal{G}_{mm'}(\omega+\epsilon'+\epsilon) .
\end{split}
\end{equation}
To further simplify the integral, we can evaluate the doubly occupied 
Green's functions at some specific value of $\omega$, as before. 
The energy dependence on $\epsilon$ for the doubly occupied Green's functions 
could be kept if we set $\omega+\epsilon'=\epsilon_{m'},$
\begin{displaymath}
\mathcal{G}_{mm'}(\omega+\epsilon'+\epsilon)\vert_{\omega+\epsilon'=\epsilon_{m'}}=\frac{-1}
{\epsilon_{m}+U-\epsilon-i\eta} .
\end{displaymath}
The final form of the vertex function is again obtained from a set of linear algebraic equations. This and the corresponding 
self-energy of the boson propagator are given, respectively, by
\begin{equation}
\Lambda_{m}^{(b)}(\omega,\epsilon)+\frac{1}{\epsilon_{m}+U-\epsilon-i\eta}
\sum_{m'\neq m}F_{m'}(\omega)\Lambda_{m'}^{(b)}(\omega,\epsilon)=1 ,
\end{equation}
and
\begin{equation}
\Sigma_{b}^{(b)}(\omega)=\sum_{m}\int\frac{d\epsilon}{\pi}~n_{F}
(\epsilon)~\Gamma_{m}(\epsilon)\Lambda_{m}^{(b)}(\omega,\epsilon)\mathcal{G}_{m}
(\omega+\epsilon) .
\end{equation}\\

To be consistent with the nomenclature introduced by Sakai, we call this approximation the 'NCA$f^2$v$\epsilon$ approximation', where $\epsilon$ indicates the double energy dependence of the vertex functions.\\

\subsection{NCA$f^2$v$\tilde\epsilon$ approximation}

The NCA$f^2$v$\epsilon$ vertex function can be further improved by evaluating the doubly occupied Green's functions (Eq. (\ref{eq:ncaf2ve})) at the value of $\omega+\epsilon'$ that most likely maximize the rest of the integrand, that is, $\omega+\epsilon' \sim \tilde{\epsilon}_{m'}$, where $\tilde\epsilon_{m'}$ are the renormalized poles of the pseudo-fermion Green's functions. 
From Eq. (\ref{eq:self-energy-fermion}), we compute directly the imaginary 
parts of the pseudo-fermion propagators, that are the spectral functions corresponding to the retarded Green's functions $\mathcal{G}_{m}(z)$
\begin{displaymath}
\rho_{m}(\omega)=-\frac{1}{\pi} Im~\mathcal{G}_{m}(\omega) .
\end{displaymath}
The poles of $\rho_{m}(\omega)$ are the energies $\tilde{\epsilon}_{m}$ that satisfy 
\begin{equation}
\tilde{\epsilon}_{m}-\epsilon_{m}-\Sigma_{m}(\tilde{\epsilon}_{m})=0 .
\end{equation}
If we set $\omega+\epsilon'=\tilde\epsilon_{m'}$ in the argument of the doubly occupied Green's functions in Eq. (\ref{eq:ncaf2ve}), the vertex functions and the boson self-energy are then given, respectively, by 

\begin{equation} \label{eq:vertex(c)}
\Lambda_{m}^{(c)}(\omega,\epsilon)+\frac{1}{\tilde{\epsilon}_{m}+U-\epsilon-i\eta}\sum_{m'\neq m}F_{m'}(\omega)\Lambda_{m'}^{(c)}(\omega,\epsilon)=1 ,
\end{equation}
\begin{equation}
\Sigma_{b}^{(c)}(\omega)=\sum_{m}\int\frac{d\epsilon}{\pi}~n_{F}
(\epsilon)~\Gamma_{m}(\epsilon)\Lambda_{m}^{(c)}(\omega,\epsilon)\mathcal{G}_{m}
(\omega+\epsilon) .
\end{equation}
\\
\subsection{Kondo temperature results}

We analyze next the effect of the different approximations described above on the Kondo temperature, within a simple model. For this purpose, we use a constant and degenerate hybrization intensity $\Gamma_{m}(\epsilon)=0.15$ eV for $-B<\epsilon<B$ and 0 otherwise. Here $B$ is the half bandwidth and we set $B=3$ eV. The degeneracy $N$ in this section is taken to be $N=6$ and we set $\epsilon_m=-2$ eV for all $m$'s.
With these parameters we solve the UNCA$^{(0)}$ set of equations in the NCA$f^2$v, NCA$f^2$v$\epsilon$, and NCA$f^2$v$\tilde\epsilon$ vertex
approximations. We obtain the spectral functions for the boson ($\rho_{b}(\omega,T)$) and pseudo-fermions ($\rho_{m}(\omega,T)$), in the $T\rightarrow 0$ limit, in order to calculate the Kondo temperature, $T_K$. As usual, this temperature is obtained from the difference between the lowest pole of $\rho_{b}(\omega)$ and the corresponding one for $\rho_{m}(\omega)$\cite{NCA-bickers-cox}. We obtain $T_K$  for different
values of the Coulomb interaction $U$ and within the different approaches to the vertex corrections previously 
introduced, taking $U$= 5 eV, 10 eV and 100 eV. We consider that $U$= 100 eV already gives the  $U\rightarrow\infty$ limit. The $U\rightarrow\infty$ limit is characterized by $\Lambda_{m}=1$ (to leading order of the large-$N$ expansion) and this limit is recovered perfectly with $U=100 $ eV. Our results are presented in Table \ref{cap:tabla-Tk}.

\begin{table}[h]
\begin{tabular*}{8 cm}{ c @{\extracolsep{\fill}} c @{\extracolsep{\fill}} c @{\extracolsep{\fill}} c }
\hline
\\
 U & $T_{K}^{(a)}$ & $T_{K}^{(b)}$ & $T_{K}^{(c)}$ \\
\\
\hline\hline
\\
 5 & 260 & 208 & 183
\\
 10 & 85 & 81 & 77
\\
100 & 33 & 34 & 31
\\
\\
\hline
\end{tabular*}
\caption{Results for the Kondo temperature obtained from different approximations to the vertex functions. The superscripts $a$, $b$ and $c$ represents the NCA$f^2$v, NCA$f^2$v$\epsilon$, and NCA$f^2$v$\tilde\epsilon$ approximations to the vertex functions respectively. The Kondo temperatures are given in Kelvin. The values of $U$ 
are given in eV. 
\label{cap:tabla-Tk}}
\end{table}

\begin{figure}[!h]
\begin{centering}\includegraphics[clip,scale=0.5]{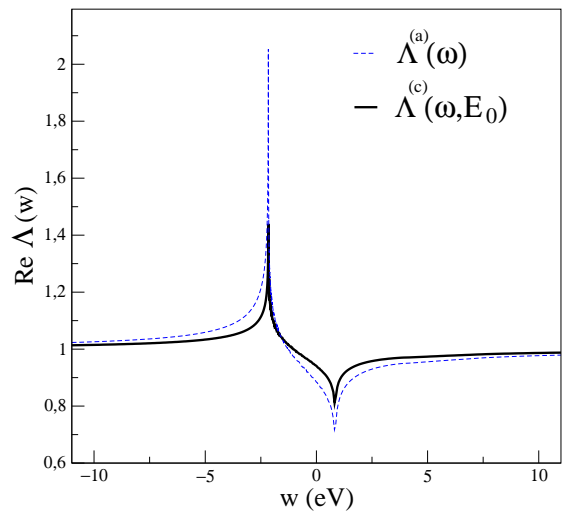}\par\end{centering}
\caption{Real part of the vertex functions as a function of $\omega$ for $U=5$ eV. 
The second argument of $\Lambda_{m}^{(c)}(\omega,\epsilon)$ is evaluated at the 
pole of the boson spectral function, $E_0$.
\label{vertex_omega}}
\end{figure}

In Fig. (\ref{vertex_omega}) we compare the real parts of the vertex functions in the NCA$f^2$v 
and NCA$f^2$v$\tilde\epsilon$ approximations as a function of $\omega$. To plot the function $Re \Lambda_{m}^{(c)}(\omega,\epsilon),$ we set its second argument equal to the value of the lowest pole 
of the boson spectral function, $E_0$. The corresponding curve for the vertex function in the 
NCA$f^2$v$\epsilon$ approximation, $\Lambda_m^{(b)}(\omega,E_0)$, is very similar to $\Lambda_{m}
^{(c)}(\omega,E_0)$, so that it is not shown in Fig. (\ref{vertex_omega}). From this figure it can be seen that the NCA$f^2$v vertex agrees qualitatively with the NCA$f^2$v$\tilde\epsilon$ one.
As it can be drawn from Table \ref{cap:tabla-Tk}, all the approximations give the same order of 
magnitude for the Kondo scale, for a given $U$. However, there is a quantitative difference in the calculated $T_K$'s that can go up to 30 $\%$.

\begin{figure}[!h]
\begin{centering}\includegraphics[clip,scale=0.5]{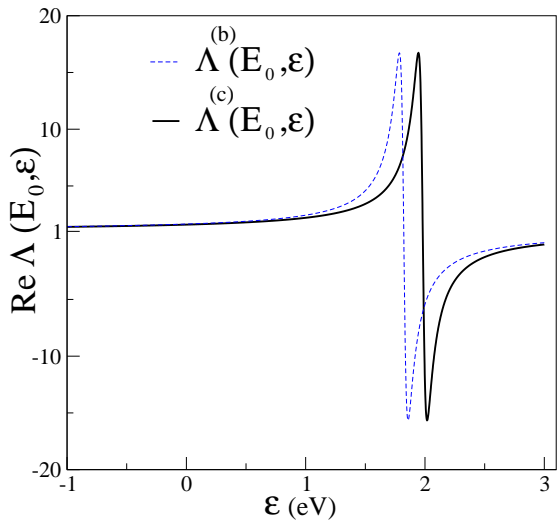}\par\end{centering}
\caption{Real part of the vertex functions in the NCA$f^2$v$\epsilon$ and NCA$f^2$v$\tilde\epsilon$ 
approximations as a function of $\epsilon$ for $U=5$ eV.
The first argument of both functions is evaluated at the pole of the boson spectral function, $E_0$.
\label{vertex_omega_epsilon}}
\end{figure}

On the other hand, in Fig. (\ref{vertex_omega_epsilon}) we plot $\Lambda_{m}^{(b)}(E_0,\epsilon)$
and $\Lambda_{m}^{(c)}(E_0,\epsilon)$ as functions of their second arguments. For $\omega$ we have chosen, just as an example the value of the boson propagator pole, $E_0$.
Both curves seem to have been rigidly shifted one with respect to the other, as a consequence of the different argument values in which the Green's functions of the 
heavy bosons are evaluated in the vertex corrections. The differences show up 
in the values obtained for $T_K$. 

Comparing Figs. (\ref{vertex_omega}) and (\ref{vertex_omega_epsilon}), it can be observed that, as a function of $\omega,$ there is an important variation of the functions in a wide energy range while, as a function of $\epsilon,$ the energy window of variation is narrower. However, in this last case the amplitude of variation is up to five times greater than in the first one. From this fact, it is clear that the second dependence in energy cannot be neglected for $U=$ 5 eV if precise values of $T_K$ are desired. This result is relevant in the case of Ce systems for which $U\sim$ 6 eV. 

\section{Crystal-field effects}

\subsection{Model}

In many real systems the crystal-field splittings are important 
and have not been taken into account in the previous model. 
In this section, we investigate a model in which the hybridization 
functions $\Gamma_{m}(\epsilon)$, defined  by 
Eq. (\ref{eq:hybridization-functions}), are energy independent, 
but have different intensities according to the corresponding 
$m$-symmetry. Therefore, in these cases there are hybridization-induced 
crystalline electric field splittings, $\Delta_{nn'}$. 
The splitting induced by hybridization is anticipated to be the 
dominant contribution to the crystal-field splittings in strongly 
hybridized Kondo systems, such as Cerium compounds\cite{Levy-Zhang}. 
The crystal-field splittings are defined as the absolute value of the difference 
between the lowest pole of the pseudo-fermion spectral functions 
($\tilde\epsilon_n$) and the excited ones ($\tilde\epsilon_{n'}$), $\Delta_{nn'}=\vert\tilde\epsilon_{n'}-\tilde\epsilon_{n}\vert$.
We consider the $\Delta_{nn'}$ coming from virtual transitions 
$f^1\rightarrow f^0\rightarrow f^1$ and 
$f^1\rightarrow f^2\rightarrow f^1$ of the local level, that is, 
we focus on the splitting coming from the anisotropy ($m$-dependence) 
of the mixing interaction $V_{km}$. 

A good scheme to solve the Hamiltonian should be able to give, not only 
the correct Kondo energy scale but also the correct symmetry 
of the local impurity ground state. 

We can get a preliminary idea about the symmetry of the ground 
state as shown in Ref. \cite{Levy-Zhang}.  
Within the infinite $U$ NCA\cite{Coleman}, the pseudo-fermions 
are dressed only by the empty boson. For a flat density of states 
of the conduction electrons, of half-width $B$, the corresponding 
real part of the self-energy of the pseudo-fermions is given, at 
zero temperature and to leading non-trivial order $1/N$, by the 
approximate relation
\begin{equation}
Re\Sigma_{m}^{U=\infty,NCA}\simeq\frac{\Gamma_{m}}{\pi}ln\Bigl\vert\frac{\epsilon_f}{B}\Bigl\vert\simeq
-\kappa\Gamma_{m},
\end{equation}
with $\kappa$ a positive real number. Therefore, the shift of the pseudo-fermion 
energies is always negative and directly proportional to the hybridization strength. 
Using this simple argument, we can foresee that the level with the strongest 
hybridization has the largest shift and becomes the ground state. It is worth to
notice that the same physics holds in the resonant model ($U=0$). 

On the other hand, within UNCA$^{(0)}$, the shift of the 
different pseudo-fermion levels is given by the self-energy coming solely 
from the doubly occupied bosons (see Eq. (\ref{eq:self-energy-fermion})). In order
to compare the $\Delta_{nn'}$'s coming from empty or doubly occupied bosons, 
we estimate this last contribution. We analyse here a system with two different constant 
hybridizations, namely $\Gamma_{1}$ and $\Gamma_{2}$,  which are  
$N_1$ and $N_2$ times degenerate, respectively. The real part of the self-energies 
of the pseudo-fermions are given by the following approximate relation, at zero 
temperature,
\begin{equation}
Re\Sigma_{m}^{U<\infty}\simeq\sum_{m'\neq m}\frac{\Gamma_{m'}}{\pi}ln\Bigl\vert\frac{\epsilon_m+U}{\epsilon_m+U+B}
\Bigl\vert\simeq-\sum_{m'\neq m}\frac{\Gamma_{m'}}{\pi}\alpha,
\end{equation}
where $\alpha$ is also a positive real number and considering that all the $\epsilon_m$ have the same value. 
We have found, taking into account only virtual transitions $f^1 \to f^0 \to f^1$ 
($U\rightarrow\infty$ limit), 
\begin{displaymath}
Re\Sigma_{1}^{U=\infty,NCA}\simeq-\kappa\Gamma_{1}
\end{displaymath}
\begin{equation}
Re\Sigma_{2}^{U=\infty,NCA}\simeq-\kappa\Gamma_{2},
\end{equation}
but on the other hand, taking into account only virtual transitions $f^1 \to f^2 \to f^1$,  ($U<\infty$), the self energies are given by 
\begin{displaymath}
Re\Sigma_{1}^{U<\infty}\simeq-\alpha\left((N_1-1)\Gamma_{1}+N_2\Gamma_{2}\right)
\end{displaymath}
\begin{equation} 
\label{1}
Re\Sigma_{2}^{U<\infty}\simeq-\alpha\left(N_1\Gamma_{1}+(N_2-1)\Gamma_{2}\right),
\end{equation}
Considering the case $\Gamma_{1}>\Gamma_{2}$, 
the corresponding self-energies are related to each other as follows
\begin{equation}
Re\Sigma_{1}^{U=\infty,NCA}<Re\Sigma_{2}^{U=\infty,NCA},
\end{equation}
\begin{equation} \label{2}
Re\Sigma_{1}^{U<\infty}>Re\Sigma_{2}^{U<\infty}.
\end{equation}
Therefore, we can infer that there exists a competition 
between both spin flip mechanisms, $f^1\rightarrow f^0\rightarrow f^1$ and
$f^1\rightarrow f^2\rightarrow f^1$, to give the ground state symmetry, as it will be confirmed in the next subsection. 

Due to this, we add to the UNCA$^{(0)}$ diagrams the fermion self-energy diagram of 
order $(1/N)^1,$ originated in virtual transitions to the empty state, and we 
analyze the influence of this last contribution on the $\Delta_{nn'}$'s values
and on the ground state symmetry.  

\begin{figure}[!h]
\begin{center}
\includegraphics[clip,scale=0.55,angle=0]{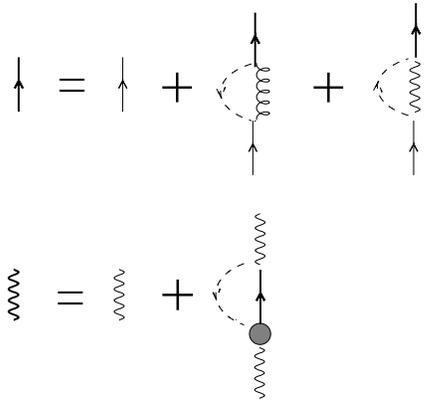}
\caption{Diagrammatic representation of the finite U non-crossing approximation up to $(1/N)^0$ order with 
an additional diagram of order $(1/N)^1$. The big dot represents the vertex function. 
The diagram of order $(1/N)^1$ has a bare light boson propagator, and it does not contain vertex corrections.
\label{cap:unca01}}
\end{center}
\end{figure}

In Fig. (\ref{cap:unca01}) we show the diagrammatic representation of the non-crossing 
approximation up to $(1/N)^0$ (UNCA$^{(0)}$) order with the additional diagram of order $(1/N)^1$. We call this approach UNCA$^{(0+1)}$ from now on. 
In order to compare on an equal footing both contributions to the self-energy 
of the pseudo fermions, in this additional diagram we neither consider vertex corrections nor the self-consistent renormalization of the light boson propagator.
Summarizing, in this approximation, the self-energy of the pseudo-fermions has two contributions, one coming from the empty boson and another from the doubly occupied ones, as shown by the following expression
\begin{displaymath} 
\Sigma_{m}{(\omega)=\sum_{m'\neq m}\int\frac{d\epsilon}{\pi}
~n_{F}(\epsilon)\Gamma_{m'}(\epsilon)\mathcal{G}_{mm'}(\omega+\epsilon)}
\end{displaymath}
\begin{equation}
\label{eq:self-energy-fermion_2}
+\int\frac{d\epsilon}{\pi}
~\left[  1-n_{F}(\epsilon)\right]  \frac{\Gamma_{m}(\epsilon)}{\omega-\epsilon}.
\end{equation}

\subsection{Crystal-field results}

As a first step in the modeling of real compounds, in this subsection we consider a system with two different constant hybridizations, $\Gamma_{1}$ and $\Gamma_{2}$,
with degeneracy $N_1=2$ and $N_2=4$, respectively. As before, we take the half bandwidth $B=3$ eV and $\epsilon_m=-2$ eV for all $m$'s. To obtain the $\Delta_{12}$'s we compute the pseudo-fermion self-energy $\Sigma_{m}(\omega)$ coming from
Eqs. (\ref{eq:self-energy-fermion}) and (\ref{eq:self-energy-fermion_2}), corresponding to UNCA$^{(0)}$ and UNCA$^{(0+1)},$ respectively, which are solved for different values of $U$. 
In this subsection we use the vertex corrections given by the NCA$f^2v\tilde\epsilon$ approximation. 
\begin{table}[h]
\begin{tabular*}{8 cm}{ c @{\extracolsep{\fill}} c @{\extracolsep{\fill}} c }
\hline
\\
$U$ & $\Delta_{12}^{UNCA^{(0)}}$ & $\Delta_{12}^{UNCA^{(0+1)}}$\\
\\
\hline\hline
\\
5 & 24 & 6

\\
10 & 12 & 21
\\
100 & 2 & 31
\\
\\
\hline
\end{tabular*}
\caption{Results for the crystal-field splitting $\Delta_{12}\equiv |\tilde{\epsilon}_1-
\tilde{\epsilon}_2|$ obtained for $\Delta\Gamma=0.1$ eV within the NCA$f^2$v$\tilde\epsilon$ approximation to the vertex functions. The values of the $\Delta_{12}$ are given in Kelvin. The values of $U$ are given in eV. 
\label{cap:tabla_UNCA0_UNCA01}}
\end{table}
\\
In Table \ref{cap:tabla_UNCA0_UNCA01} we show the calculated values of $\Delta_{12}$ for $\Gamma_1=0.15$ eV and $\Gamma_2=0.05$ eV. We obtain the $\Gamma_2$ symmetry as the ground state when we 
use UNCA$^{(0)}$ and a decrease of the absolute value of $\Delta_{12}$ 
when going from $U=5$ eV to $U\rightarrow\infty$. It is clear from Eq. (\ref{1}) that
$\alpha$ vanishes in the $U\rightarrow\infty$ limit and that the shift produced by $\Sigma_{1}^{U<\infty}$
becomes equal to the one corresponding to $\Sigma_{2}^{U<\infty}$. The crystal-field splittings in the 
infinite $U$ limit come only from the $f^1\rightarrow f^0\rightarrow f^1$ processes, as expected.

On the other hand, UNCA$^{(0+1)}$ gives $\Gamma_1$ as the ground state symmetry, as it has been anticipated by the approximate expression of Eq. (\ref{2}). In this case, we obtain an increase in the values of $\Delta_{12}$ when 
going from $U=5$ to $U\rightarrow\infty$. Similar behaviour is obtained for different values of $\Delta\Gamma=\Gamma_1-\Gamma_2$.
\begin{figure}[!h]
\begin{center}
\includegraphics[clip,scale=0.5,angle=0]{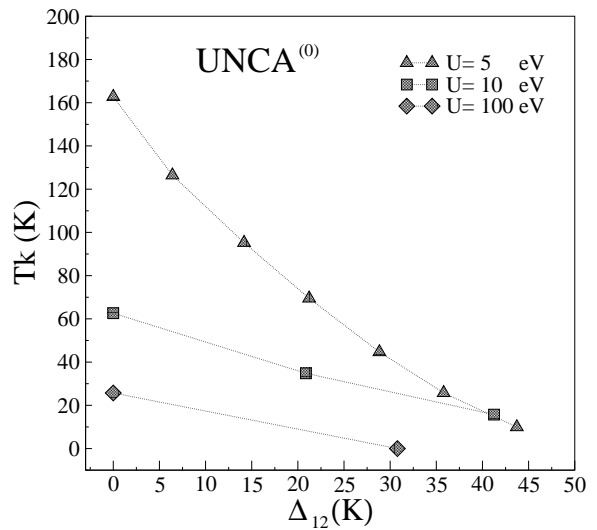}
\caption{Kondo temperature, $T_K,$ as a function of the  crystal-field splitting, $\Delta_{12}$, 
within the UNCA$^{(0)}$ scheme. The obtained symmetry for the ground state is $\Gamma_{2}$ in all 
calculations. The values are given in Kelvin.\label{Tk_CEF_unca0}}
\end{center}
\end{figure}

\begin{figure}[!h]
\begin{center}
\includegraphics[clip,scale=0.5,angle=0]{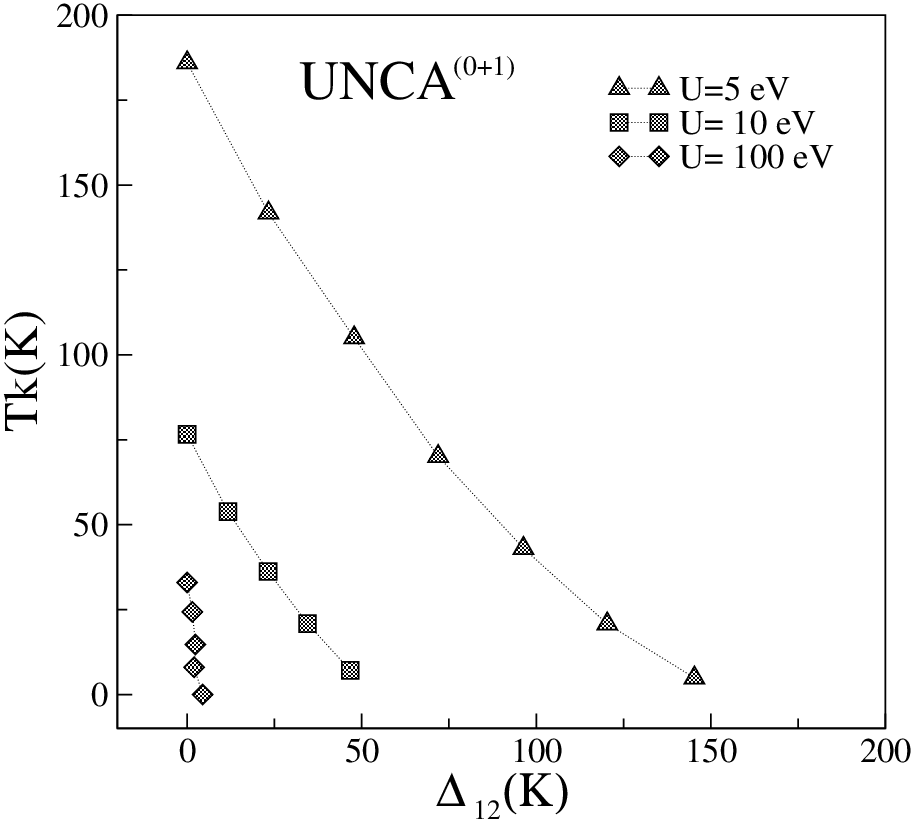}
\caption{Kondo temperature, $T_K,$ as a funcions of the  crystal-field splittings, $\Delta_{12}$, 
within the UNCA$^{(0+1)}$ scheme. The obtained symmetry for the ground state is $\Gamma_{1}$ in all 
calculations. The values are given in Kelvin.\label{Tk_CEF_unca01}}
\end{center}
\end{figure}

We then analyse the evolution of $T_K$ for different values of $\Delta\Gamma$ by setting $\Gamma_{1}=0.15$ eV and varying  $\Gamma_{2},$ from 0.09 to 0.15 eV.  
 
For each $\Delta\Gamma$ we calculate the splitting $\Delta_{12}$ and $T_K$ within UNCA$^{(0)}$ and UNCA$^{(0+1)}$ as shown in Figs. (\ref{Tk_CEF_unca0}) and (\ref{Tk_CEF_unca01}) respectively. In both 
cases, and for all the values of $U$ considered, we obtain a decrease of the Kondo temperature as the
crystal field splitting increases. 
In this simple context using constant hybridization functions, the increment in the splitting of the levels reduces the 
effective degeneracy of the ground state which, in turn, is correlated with a decreasing  $T_K$.
This behaviour is more pronounced when using UNCA$^{(0+1)}$. 

On the other hand, it has to be noticed that in the UNCA$^{(0)}$ calculation, when  $U$ is large,
a small variation in $\Delta\Gamma$ gives rise to large variations in $\Delta_{12}$. This leads to low Kondo temperatures and even to the disappearence of the Kondo regime ($T_K<0$) which can be clearly seen in Fig. (\ref{Tk_CEF_unca0}) for $U$= 100 eV, where only two points present a positive $T_K$.

It must be stressed that the evolution of $T_K$ with $\Delta_{12}$ shown in Fig. (\ref{Tk_CEF_unca0}) and (\ref{Tk_CEF_unca01}) is not necessarily the one of a real compound. We will show in the next section that using \emph{ab initio} calculated hybridization functions, which show a rich structure as a function of energy, can give rise to a different behavior.
\\

\section{Realistic calculations for Cerium compounds}

In this section we apply the previous ideas to two realistic Cerium 
compounds with different electronic properties. In the case of Ce intermetallic compounds long experience in the area shows that the behavior of different properties, such as T$_K$, crystal field splittings, magnetic susceptibilities,  can be readily obtained, qualitatively as well as quantitatively, by considering an impurity problem due to the localization of the 4f orbitals \cite{Han,Svane1,Vero-5,paper-CeXY,Ehm,Moreno}.
In particular, we choose 
the heavy fermion system CeIn$_3$ and the intermediate valence one 
CeSn$_3$. In both systems, from magnetic susceptibility measurements, 
Pedrazzini \emph{et al} \cite{Pedrazzini} have obtained the 
$J=\frac{5}{2}$ $\Gamma_7$ doublet as the ground state. The same result was 
obtained from neutron scattering measurements by Murani \emph{et al} 
\cite{Murani}. CeIn$_3$ shows the normal Curie-Weiss susceptibility at 
all temperatures except around and below the N\'eel temperature, $T_N(=10.2 K)$, 
so that a small value for the Kondo temperature is expected ($T_K<T_N$) \cite{Murani}. 
CeSn$_3$, on the other hand, shows an enhanced Pauli susceptibility  at 
low temperatures followed by a shallow maximum and a Curie-Weiss behaviour above it \cite{Murani}. The Kondo temperature in this case is well reported and its value is
$T_K=450 K$\cite{Han}.
 
In both compounds a precise experimental estimation of the crystal-field splitting is difficult. In the case of CeIn$_3$, In is a strong neutron absorber
and this  makes the interpretations of the experiments rather cumbersome. In spite of this, the reported estimated value for $\Delta_{CF}$ is around 130 K \cite{Murani,Pedrazzini}. On the other hand, in the case of CeSn$_3$ the strong hybridization induces a large $T_K$ so that $T_K >> \Delta_{CF}$ and the experiments fail to resolve the crystal-field peaks. 

\subsection{\emph{Ab initio} hybridization functions}

We obtain the hybridization functions of the 4f states of Cerium with 
the conduction band from first principles within the Density Functional 
Theory. In this work, the \emph{ab initio} calculations are done using 
the full potential linearized augmented plane waves method (FP-LAPW), 
as implemented in the Wien2k code \cite{Wien}. As suggested by 
Gunnarsson \emph{et al.} \cite{Gunnarsson-hib}, the hybridization $\Gamma_{m}(\varepsilon)$ can be estimated 
from the projected LDA 4f density matrix $\rho_{m}^{LDA}$ at the Ce 
site in the following way, 

\begin{equation}
\Gamma_{m}(\varepsilon)=-Im\left\{
\lim_{\eta\rightarrow0}\left[\left(\int
dz\frac{\rho_{m}^{LDA}(z)}{\varepsilon-i\eta-z}\right)\right]^{-1}\right\}.
\label{eq:hib}
\end{equation}

In all cases the labels \emph{m} correspond to the different irreducible 
representations of the 4f states at the cubic Ce site. That is, for 
$J=\frac{5}{2}$ the doublet $\Gamma_7$ and the quartet $\Gamma_8$, 
while for $J=\frac{7}{2}$ the doublets $\Gamma_6$ and $\Gamma_7,$ and 
the quartet $\Gamma_8$. The LDA  calculations are performed  at the 
experimental volumes of the CeIn$_3$ and CeSn$_3$ compounds. The 
\textit{muffin-tin} radii, $R_{mt}$ are taken equal to 2.4 a.u. in the 
case of the anion ligands, while the corresponding radii for Ce are taken 
equal to 3.0 a.u. in the CeIn$_3$ compound and 3.3 a.u. in the CeSn$_3$ 
one. 102 \textbf{k} points in the irreducible Brillouin zone are considered 
to be enough  for the quantities to be calculated.

The hybridization function is used as input in the UNCA set of equations. 
The crystal-field splittings are read from the separation of the peaks of 
the different spectral functions, $\rho_{m}$'s, which are shifted one 
with respect to the other due to the different degree of hybridization of 
each 4$f$ level with the conduction band. We focus on the value of the 
splitting in the $J=\frac{5}{2}$ multiplet, namely  $\Delta_{CF}=\varepsilon_{f\Gamma_7}-\varepsilon_{f\Gamma_8}$. We will 
call this mixed technique LDA-UNCA from now on. In the UNCA equations 
we take the bare energy value for the 4$f$ state from photoemission 
experiments \cite{Allen}. It is, namely, -2 eV for CeIn$_3$ and 
CeSn$_3$. We shift the $J=\frac{7}{2}$ multiplet in an amount 
given by $\Delta_{SO}=0.35~eV$, due to the spin-orbit interaction,
and use $U=$6 eV for the on-site Coulomb interaction constant among 
the $f$-electrons, which is considered to be a standard value for Cerium 
systems \cite{constrained-LDA}.
All these energy levels are given with respect to the Fermi energy.

\begin{figure}[h]
\begin{centering}
\includegraphics[clip,scale=0.5]{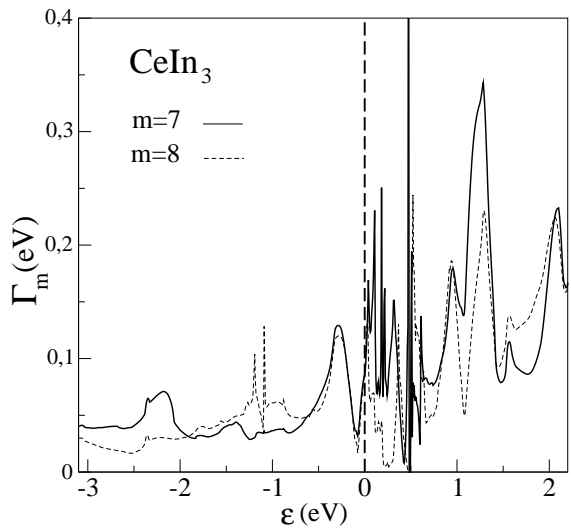}\par\end{centering}
\caption{Hybridizations functions from LDA calculations for the $J=5/2$ 
multiplet of CeIn$_3$.
\label{CeIn3_hyb}}
\end{figure}
\begin{figure}[h]
\begin{centering}
\includegraphics[clip,scale=0.5]{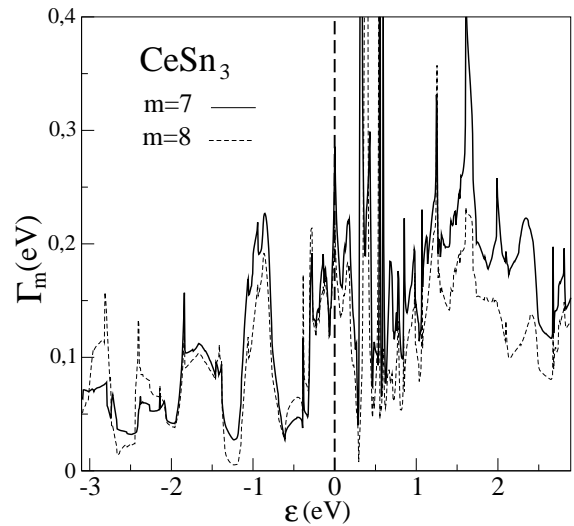}\par\end{centering}
\caption{Hybridizations functions from LDA calculations for the $J=5/2$ 
multiplet of CeSn$_3$.
\label{CeSn3_hyb}}
\end{figure}

\subsection{LDA-UNCA results}

The calculated hybridization functions for the $J=5/2$ multiplet are shown 
in Figs. (\ref{CeIn3_hyb}) and (\ref{CeSn3_hyb}) in the cases of CeIn$_3$ 
and CeSn$_3,$ respectively. These hybridization functions are in qualitative agreement 
with the ones calculated by Han \emph{et al.} \cite{Han} using the LMTO-ASA approximation. 
In both compounds, the $\Gamma_{7}$ symmetry 
has the largest average value in the whole range of energy. Using the
LDA-UNCA method, we obtain a ground state of $\Gamma_{8}$ symmetry 
for both systems when we use the UNCA$^{(0)}$ scheme. 
This does not agree with the experimental results for these Cerium compounds. 
The $\Delta_{CF}$'s do not depend on the approximation employed to compute 
the vertex functions and their values are $\Delta_{CF}=18 K$ and 
$\Delta_{CF}=60 K$ for CeIn$_3$ and CeSn$_3,$ respectively. 
On the other hand, within the UNCA$^{(0+1)}$ scheme, we obtain a  
ground state with $\Gamma_{7}$ symmetry for both studied Cerium systems, 
and for all vertex corrections used. In this scheme, the values of the 
$\Delta_{CF}$ are 150 K and 180 K for CeIn$_3$ and CeSn$_3$, respectively. In the first case this is 
in good agreement with the experimental result reported for CeIn$_3$ (130 K). 
As we have shown in Section III D, the $T_K$ strongly depend 
on the choice of the approximation for the vertex functions. 
For these real systems, the $T_K$ are again overestimated in 
the NCA$f^2$v approximation with respect to the NCA$f^2$v$\tilde\epsilon$ 
one. 
In Table \ref{cap:tabla-Tk-CeSn3} we show the calculated values for the $T_K$, 
as well as the symmetry of the ground state, for the CeSn$_3$ 
compound in both, UNCA$^{(0)}$ and UNCA$^{(0+1)}$ schemes, and for all the 
approximated vertices considered.

We want to remark that the UNCA$^{(0+1)}$ scheme with the 
NCA$f^2$v$\tilde\epsilon$ approximation for the vertex corrections gives 
 values for the ground state symmetry and for 
the $\Delta_{CF}$'s in very good agreement with the experimental ones in these Cerium systems. 
Furthermore, for the CeSn$_3$ compound, we obtain a value for $T_K$ very near to the experimental one, 
that is,$450 K$. In the case of CeIn$_3$  we obtain a negative value 
for the $T_K$ which implies that the Kondo regime has not been completely established in agreement again with the experimental observation that $T_K$, if it exists, should be below $T_N\sim10$ K as mentioned above.

All the vertex approximations presented in this work capture 
the experimental energy scale of $T_K$ in the case of CeSn$_3,$ but only the NCA$f^2$v$\tilde\epsilon$ approximation within the UNCA$^{(0+1)}$
scheme gives a quantitative agreement ($T_K=480 K$) with the experimental 
value ($T_K=450 K$).

\begin{table}[h]
\begin{tabular*}{8 cm}{c@{\extracolsep{\fill}}c@{\extracolsep{\fill}}c@{\extracolsep{\fill}}c@{\extracolsep{\fill}}c}
\hline
\\
CeSn$_3$ & $\Gamma_{GS}$ & NCA$f^2$v & NCA$f^2$v$\epsilon$ & NCA$f^2$v$\tilde\epsilon$
\\
 & & $T_K$ & $T_K$ & $T_K$
\\
\hline\hline
\\
UNCA$^{(0)}$ & $\Gamma_{8}$ & 1030 & 788 & 722
\\
\\
UNCA$^{(0+1)}$ & $\Gamma_{7}$ & 757 & 538 & 480
\\
\\
\hline
\end{tabular*}
\caption{Results for the Kondo temperature and symmetry of the ground state ($\Gamma_{GS}$) as obtained using different approximations to the vertex function. Within the UNCA$^{(0)}$ and UNCA$^{(0+1)}$ schemes for the CeSn$_3$ compound. 
The Kondo temperatures are given in Kelvin.\label{cap:tabla-Tk-CeSn3}}
\end{table}
As already mentioned, the results obtained for $T_K$ and for the crystal-field splittings in these compounds do not follow the behavior of the constant hybridization simple model of the previous section, in the sense that a larger splitting does not necessarily lead to a smaller $T_K$. This  enforces the importance of taking into account the real electronic structure information when solving many-body hamiltonians.

\section{Discussion and conclusions}

We have solved the Anderson impurity Hamiltonian using the finite $U$ non-crossing approximation (UNCA) in its simplest scheme, that is, considering the self energies at lowest order in the $1/N$ diagrammatic expansion (UNCA$^{(0)}$). This approximation yields a correct Kondo energy scale, and it is then a reliable approach 
to perform realistic calculations for 4f systems. 

Even if the approximation is simple, a set of coupled integral equations for the vertex functions has to be 
solved. These functions depend on two independent energy variables. 
In this work, we have extended the NCA$f^2$v approximation to the vertex functions introduced by Sakai \textit{et al.}, including the second energy variable in two different ways called NCA$f^2$v$\epsilon$ and  NCA$f^2$v$\tilde\epsilon$.
We have calculated and compared the T$_K$ using these three different approximations. Our results confirm that the NCA$f^2$v approximation gives the correct Kondo energy scale,
but, in order to make a quantitative comparison with the experiments (in real systems), the double dependence on energy cannot be disregarded when the Coulomb interaction involved is near to $U=6$ eV. In fact, with our NCA$f^2$v$\tilde\epsilon$ approximation, we obtain T$_K$ for the CeSn$_3$ compound in surprising
agreement with the experimental one, when using \emph{ab initio} calculated hybridization functions.

We have also analyzed the competition between the two spin flip mechanisms, involving virtual transitions 
to empty and doubly occupied states, in the determination of the ground state symmetry of the local state, 
by including an extra diagram of higher order in $1/N$. This extra diagram corresponds physically to virtual transitions to the empty state, $f^1\rightarrow f^0\rightarrow f^1$, which is not included in the UNCA$^{(0)}$ scheme. We have shown that the inclusion of this diagram (UNCA$^{(0+1)}$) is necessary
in order to obtain the experimental ground state symmetry for the Cerium systems CeIn$_3$ and CeSn$_3$, 
which we have taken as test examples.

Summarizing, this work presents an analysis of the different corrections that can be made on the vertex functions within the finite-U NCA approximation showing, quantitatively, the importance of its double energy dependence on the Kondo temperature. It also gives physical insight into the factors  determining the ground state symmetries as a competition between two different spin flip mechanisms. We finally test and apply the previous features in real systems taking as input the LDA hybridizations, obtaining very good agreement with experimental information.

\section{appendix}

\subsection{Numerical Details}
All vertex corrections and self-energies have been computed directly from their expressions 
given in this paper. To calculate them, we employ two different sets of frequencies, $\epsilon(i)$ 
and $\omega(j)$ with $1\leq i \leq M_1$ and $1\leq j \leq M_2$ respectively. 
The mesh for $\epsilon(i)$, in which the hybridization function is defined, is divided into $M_1$ equal steps of size $h=(e_{max}-e_{min})/M_1$. We use a large value for $M_1,$ typically 4000. In view of this, a simple trapezoid rule is enough to perform the 
integrals. On the other hand, the mesh $\omega(j)$ is a non-linear one and collects the majority of its points near the pole of the light boson spectral function. This mesh is also dense enough near the pole of the pseudo-fermions spectral functions. This attribute becomes fundamental to give a  well defined $T_K$ working with a reasonable amount of $w$ frequencies. In this analysis, we use a number $M_2$ of $\omega(j)$ frequencies around 1000-1500. The infinitesimal $\eta$ in the Green's functions is fixed as a multiple of the minor step in the  $\omega(j)$ mesh.

\subsection{Linear algebraic equations for $\Lambda_{m}^{(c)}(\omega,\epsilon)$}
In this subsection, we give some details about the set of linear algebraic equations 
given by Eq. (\ref{eq:vertex(c)}). We use the short notation 
\begin{displaymath}
\varepsilon_{m}(\epsilon)=\tilde{\epsilon}_{m}+U-\epsilon-i\eta
\end{displaymath}
for the equation that defines the vertex $\Lambda_{m}^{(c)}$, which becomes 
\begin{displaymath}
\varepsilon_{m}(\epsilon)\Lambda_{m}^{(c)}(\omega,\epsilon)+\sum_{m'\neq m}F_{m'}(\omega)\Lambda_{m'}^{(c)}(\omega,\epsilon)=\varepsilon_{m}(\epsilon)
\end{displaymath}
The functions $F_{m}(\omega)$ are given by Eq. (\ref{eq:F_{m}}).

In the specific case in which the index $m$ runs over three different symmetries,
namely $[1,2,3],$ whose degeneracies are $\left\lbrace 2,4,8\right\rbrace,$ respectively, 
the solution of the previous system requires the inversion 
of a complex non-symmetrical $3\times 3$ matrix. This inversion is done for each 
$\omega(j)$ and $\epsilon(i)$ energies. \begin{displaymath}
\Lambda^{(c)}(\omega,\epsilon)=\textbf{M}^{-1}(\omega,\epsilon)~\varepsilon(\epsilon)
\end{displaymath}
where the matrix $\textbf{M}$, the vectors $\Lambda^{(c)}(\omega,\epsilon)$ and $\varepsilon(\epsilon)$
are given by the following expressions
\begin{displaymath}
\textbf{M}(\omega,\epsilon)=\left( \begin{array}{ccc}
F_1(\omega) + \varepsilon_{1}(\epsilon) & 4F_2(\omega) &  8F_3(\omega)  \\
2F_1(\omega)  & 3F_2(\omega)+ \varepsilon_{2}(\epsilon) &  8F_3(\omega) \\
2F_1(\omega)  & 4F_2(\omega) &  7F_3(\omega)+ \varepsilon_{3}(\epsilon)
\end{array} \right )
\end{displaymath}

\begin{displaymath}
\Lambda^{(c)}(\omega,\epsilon)=\left( \begin{array}{c}
\Lambda_{1}^{(c)}(\omega,\epsilon) \\
\Lambda_{2}^{(c)}(\omega,\epsilon) \\
\Lambda_{3}^{(c)}(\omega,\epsilon)
\end{array} \right )
\end{displaymath}
\\
\begin{displaymath}
\varepsilon=\left( \begin{array}{c}
\varepsilon_{1}(\epsilon) \\
\varepsilon_{2}(\epsilon) \\
\varepsilon_{3}(\epsilon)
\end{array} \right )
\end{displaymath}

This is the typical case of Cerium systems in which level $m=1$ and $m=2$ correspond to 
the $\Gamma_7$, $\Gamma_8$ symmetries of $j=5/2$ multiplet, respectively, and $m=3$ 
represents the whole $j=7/2$ multiplet.\\

\section{ACKNOWLEDGMENTS}

This work was partially funded by Fundaci\'on YPF, CNEA, UBACyT-X115,
PICT-0310698 and PIP
2005-2006 Num. 6016. A. M. Llois, V. L.
Vildosola and L.O. Manuel belong to CONICET (Argentina).
\bibliographystyle{apsrev}
\bibliography{tesis}

\begin{thebibliography}{27}
\expandafter\ifx\csname natexlab\endcsname\relax\def\natexlab#1{#1}\fi
\expandafter\ifx\csname bibnamefont\endcsname\relax
  \def\bibnamefont#1{#1}\fi
\expandafter\ifx\csname bibfnamefont\endcsname\relax
  \def\bibfnamefont#1{#1}\fi
\expandafter\ifx\csname citenamefont\endcsname\relax
  \def\citenamefont#1{#1}\fi
\expandafter\ifx\csname url\endcsname\relax
  \def\url#1{\texttt{#1}}\fi
\expandafter\ifx\csname urlprefix\endcsname\relax\def\urlprefix{URL }\fi
\providecommand{\bibinfo}[2]{#2}
\providecommand{\eprint}[2][]{\url{#2}}

\bibitem[{\citenamefont{{N. E. Bickers} et~al.}(1987)\citenamefont{{N. E.
  Bickers}, {D. L. Cox}, and {J. W. Wilkins}}}]{NCA-bickers-cox}
\bibinfo{author}{\bibnamefont{{N. E. Bickers}}},
  \bibinfo{author}{\bibnamefont{{D. L. Cox}}}, \bibnamefont{and}
  \bibinfo{author}{\bibnamefont{{J. W. Wilkins}}}, \bibinfo{journal}{Phys. Rev.
  B} \textbf{\bibinfo{volume}{36}}, \bibinfo{pages}{2036}
  (\bibinfo{year}{1987}).

\bibitem[{\citenamefont{{N. E. Bickers}}(1987)}]{Bickers}
\bibinfo{author}{\bibnamefont{{N. E. Bickers}}}, \bibinfo{journal}{Rev. Mod.
  Phys.} \textbf{\bibinfo{volume}{59}}, \bibinfo{pages}{845}
  (\bibinfo{year}{1987}).

\bibitem[{\citenamefont{{P. Coleman}}(1984)}]{Coleman}
\bibinfo{author}{\bibnamefont{{P. Coleman}}}, \bibinfo{journal}{Phys. Rev. B}
  \textbf{\bibinfo{volume}{29}}, \bibinfo{pages}{3035} (\bibinfo{year}{1984}).

\bibitem[{\citenamefont{{J. E. Han} et~al.}(1997)\citenamefont{{J. E. Han}, {M.
  Alouani}, and {D. L. Cox}}}]{Han}
\bibinfo{author}{\bibnamefont{{J. E. Han}}}, \bibinfo{author}{\bibnamefont{{M.
  Alouani}}}, \bibnamefont{and} \bibinfo{author}{\bibnamefont{{D. L. Cox}}},
  \bibinfo{journal}{Phys. Rev. Lett.} \textbf{\bibinfo{volume}{78}},
  \bibinfo{pages}{939} (\bibinfo{year}{1997}).

\bibitem[{\citenamefont{{J. Laegsgaard} and {A.
  Svane}}(1998{\natexlab{a}})}]{Svane1}
\bibinfo{author}{\bibnamefont{{J. Laegsgaard}}} \bibnamefont{and}
  \bibinfo{author}{\bibnamefont{{A. Svane}}}, \bibinfo{journal}{Phys. Rev. B}
  \textbf{\bibinfo{volume}{58}}, \bibinfo{pages}{12817}
  (\bibinfo{year}{1998}{\natexlab{a}}).

\bibitem[{\citenamefont{{P. Roura-Bas} et~al.}(2007)\citenamefont{{P.
  Roura-Bas}, {V. Vildosola}, and {A. M. Llois}}}]{paper-CeXY}
\bibinfo{author}{\bibnamefont{{P. Roura-Bas}}},
  \bibinfo{author}{\bibnamefont{{V. Vildosola}}}, \bibnamefont{and}
  \bibinfo{author}{\bibnamefont{{A. M. Llois}}}, \bibinfo{journal}{Phys. Rev.
  B} \textbf{\bibinfo{volume}{75}}, \bibinfo{pages}{195129}
  (\bibinfo{year}{2007}).

\bibitem[{\citenamefont{{R. Aguado} and {D. Langreth}}(2003)}]{Aguado-Langreth}
\bibinfo{author}{\bibnamefont{{R. Aguado}}} \bibnamefont{and}
  \bibinfo{author}{\bibnamefont{{D. Langreth}}}, \bibinfo{journal}{Phys. Rev.
  B} \textbf{\bibinfo{volume}{67}}, \bibinfo{pages}{245307}
  (\bibinfo{year}{2003}).

\bibitem[{\citenamefont{{N.S. Wingreen} and {Y.M. Meir}}(1994)}]{Wingreen}
\bibinfo{author}{\bibnamefont{{N.S. Wingreen}}} \bibnamefont{and}
  \bibinfo{author}{\bibnamefont{{Y.M. Meir}}}, \bibinfo{journal}{Phys. Rev. B}
  \textbf{\bibinfo{volume}{49}}, \bibinfo{pages}{11040} (\bibinfo{year}{1994}).

\bibitem[{\citenamefont{{Th. Pruschke} and {N. Grewe}}(1989)}]{Pruschke89}
\bibinfo{author}{\bibnamefont{{Th. Pruschke}}} \bibnamefont{and}
  \bibinfo{author}{\bibnamefont{{N. Grewe}}}, \bibinfo{journal}{Z. Phys. B}
  \textbf{\bibinfo{volume}{74}}, \bibinfo{pages}{439} (\bibinfo{year}{1989}).

\bibitem[{\citenamefont{{O. Sakai} et~al.}(2005)\citenamefont{{O. Sakai}, {Y.
  Shimizu}, and {Y. Kaneta}}}]{sakay}
\bibinfo{author}{\bibnamefont{{O. Sakai}}}, \bibinfo{author}{\bibnamefont{{Y.
  Shimizu}}}, \bibnamefont{and} \bibinfo{author}{\bibnamefont{{Y. Kaneta}}},
  \bibinfo{journal}{J. Phys. Soc. Jap.} \textbf{\bibinfo{volume}{74}},
  \bibinfo{pages}{2517} (\bibinfo{year}{2005}).

\bibitem[{\citenamefont{{J. Otsuki} and {Y. Kuramoto}}(2006)}]{Otsuki}
\bibinfo{author}{\bibnamefont{{J. Otsuki}}} \bibnamefont{and}
  \bibinfo{author}{\bibnamefont{{Y. Kuramoto}}}, \bibinfo{journal}{J. Phys.
  Soc. Jap.} \textbf{\bibinfo{volume}{75}}, \bibinfo{pages}{064707}
  (\bibinfo{year}{2006}).

\bibitem[{\citenamefont{{K. Haule } et~al.}(2001)\citenamefont{{K. Haule }, {S.
  Kirchner}, {J. Kroha}, and {P. W\"olfle}}}]{Haule}
\bibinfo{author}{\bibnamefont{{K. Haule }}}, \bibinfo{author}{\bibnamefont{{S.
  Kirchner}}}, \bibinfo{author}{\bibnamefont{{J. Kroha}}}, \bibnamefont{and}
  \bibinfo{author}{\bibnamefont{{P. W\"olfle}}}, \bibinfo{journal}{Phys. Rev.
  B} \textbf{\bibinfo{volume}{64}}, \bibinfo{pages}{155111}
  (\bibinfo{year}{2001}).

\bibitem[{\citenamefont{{K. Kang} and {B.I. Min}}(1996)}]{Kang-Min}
\bibinfo{author}{\bibnamefont{{K. Kang}}} \bibnamefont{and}
  \bibinfo{author}{\bibnamefont{{B.I. Min}}}, \bibinfo{journal}{Phys. Rev. B}
  \textbf{\bibinfo{volume}{54}}, \bibinfo{pages}{1645} (\bibinfo{year}{1996}).

\bibitem[{\citenamefont{{T.A. Costi} et~al.}(1996)\citenamefont{{T.A. Costi},
  {J. Kroha}, and {P. W\"olfle}}}]{Costi_Kroha_Wolfle}
\bibinfo{author}{\bibnamefont{{T.A. Costi}}}, \bibinfo{author}{\bibnamefont{{J.
  Kroha}}}, \bibnamefont{and} \bibinfo{author}{\bibnamefont{{P. W\"olfle}}},
  \bibinfo{journal}{Phys. Rev. B} \textbf{\bibinfo{volume}{53}},
  \bibinfo{pages}{1850} (\bibinfo{year}{1996}).

\bibitem[{\citenamefont{{L. P. Kadanoff} and {G. Baym}}(1961)}]{kadanoff-baym}
\bibinfo{author}{\bibnamefont{{L. P. Kadanoff}}} \bibnamefont{and}
  \bibinfo{author}{\bibnamefont{{G. Baym}}}, \bibinfo{journal}{Phys. Rev.}
  \textbf{\bibinfo{volume}{124}}, \bibinfo{pages}{287} (\bibinfo{year}{1961}).

\bibitem[{\citenamefont{{G. Baym}}(1962)}]{baym}
\bibinfo{author}{\bibnamefont{{G. Baym}}}, \bibinfo{journal}{Phys. Rev.}
  \textbf{\bibinfo{volume}{127}}, \bibinfo{pages}{1391} (\bibinfo{year}{1962}).

\bibitem[{\citenamefont{{J. Laegsgaard} and {A.
  Svane}}(1998{\natexlab{b}})}]{Svane}
\bibinfo{author}{\bibnamefont{{J. Laegsgaard}}} \bibnamefont{and}
  \bibinfo{author}{\bibnamefont{{A. Svane}}}, \bibinfo{journal}{Phys. Rev. B}
  \textbf{\bibinfo{volume}{58}}, \bibinfo{pages}{12817}
  (\bibinfo{year}{1998}{\natexlab{b}}).

\bibitem[{\citenamefont{{P. M. Levy} and {S. Zhang}}(1989)}]{Levy-Zhang}
\bibinfo{author}{\bibnamefont{{P. M. Levy}}} \bibnamefont{and}
  \bibinfo{author}{\bibnamefont{{S. Zhang}}}, \bibinfo{journal}{Phys. Rev.
  Lett.} \textbf{\bibinfo{volume}{62}}, \bibinfo{pages}{78}
  (\bibinfo{year}{1989}).

\bibitem[{\citenamefont{{V. L. Vildosola} et~al.}(2005)\citenamefont{{V. L.
  Vildosola}, {A. M. Llois}, and {M. Alouani}}}]{Vero-5}
\bibinfo{author}{\bibnamefont{{V. L. Vildosola}}},
  \bibinfo{author}{\bibnamefont{{A. M. Llois}}}, \bibnamefont{and}
  \bibinfo{author}{\bibnamefont{{M. Alouani}}}, \bibinfo{journal}{Phys. Rev. B}
  \textbf{\bibinfo{volume}{71}}, \bibinfo{pages}{184420}
  (\bibinfo{year}{2005}).

\bibitem[{\citenamefont{{D. Ehm} et~al.}(2007)\citenamefont{{D. Ehm}, {S.
  H\"ufner}, {F. Reinert}, {J. Kroha}, {P.W\"olfle}, {O. Stockert}, {C.
  Geibel}, and {H. v. L\"ohneysen}}}]{Ehm}
\bibinfo{author}{\bibnamefont{{D. Ehm}}}, \bibinfo{author}{\bibnamefont{{S.
  H\"ufner}}}, \bibinfo{author}{\bibnamefont{{F. Reinert}}},
  \bibinfo{author}{\bibnamefont{{J. Kroha}}},
  \bibinfo{author}{\bibnamefont{{P.W\"olfle}}},
  \bibinfo{author}{\bibnamefont{{O. Stockert}}},
  \bibinfo{author}{\bibnamefont{{C. Geibel}}}, \bibnamefont{and}
  \bibinfo{author}{\bibnamefont{{H. v. L\"ohneysen}}}, \bibinfo{journal}{Phys.
  Rev. B} \textbf{\bibinfo{volume}{76}}, \bibinfo{pages}{045117}
  (\bibinfo{year}{2007}).

\bibitem[{\citenamefont{{N.O.Moreno} et~al.}(2005)\citenamefont{{N.O.Moreno},
  {A. Lobos}, {A.A. Aligia}, {E. D. Bauer}, {S. Bobev}, {V. Fritsch}, {J. L.
  Sarrao}, {P. G. Pagliuso}, {J. D. Thompson}, {C. D. Batista}
  et~al.}}]{Moreno}
\bibinfo{author}{\bibnamefont{{N.O.Moreno}}}, \bibinfo{author}{\bibnamefont{{A.
  Lobos}}}, \bibinfo{author}{\bibnamefont{{A.A. Aligia}}},
  \bibinfo{author}{\bibnamefont{{E. D. Bauer}}},
  \bibinfo{author}{\bibnamefont{{S. Bobev}}}, \bibinfo{author}{\bibnamefont{{V.
  Fritsch}}}, \bibinfo{author}{\bibnamefont{{J. L. Sarrao}}},
  \bibinfo{author}{\bibnamefont{{P. G. Pagliuso}}},
  \bibinfo{author}{\bibnamefont{{J. D. Thompson}}},
  \bibinfo{author}{\bibnamefont{{C. D. Batista}}}, \bibnamefont{et~al.},
  \bibinfo{journal}{Phys. Rev. B} \textbf{\bibinfo{volume}{71}},
  \bibinfo{pages}{165107} (\bibinfo{year}{2005}).

\bibitem[{\citenamefont{{P. Pedrazzini \emph{et al.}}}(2001)}]{Pedrazzini}
\bibinfo{author}{\bibnamefont{{P. Pedrazzini \emph{et al.}}}},
  \bibinfo{journal}{J. Magn. Magn. Mater.} \textbf{\bibinfo{volume}{226-230}},
  \bibinfo{pages}{161} (\bibinfo{year}{2001}).

\bibitem[{\citenamefont{{A.P. Murani} et~al.}(1993)\citenamefont{{A.P. Murani},
  {A.D. Taylor}, {R. Osborn}, and {Z.A. Bowden}}}]{Murani}
\bibinfo{author}{\bibnamefont{{A.P. Murani}}},
  \bibinfo{author}{\bibnamefont{{A.D. Taylor}}},
  \bibinfo{author}{\bibnamefont{{R. Osborn}}}, \bibnamefont{and}
  \bibinfo{author}{\bibnamefont{{Z.A. Bowden}}}, \bibinfo{journal}{Physical
  Review B} \textbf{\bibinfo{volume}{48}}, \bibinfo{pages}{10606}
  (\bibinfo{year}{1993}).

\bibitem[{\citenamefont{{P. Blaha} et~al.}(1999)\citenamefont{{P. Blaha}, {K.
  Schwarz}, {G.Madsen}, {D.Kvasnicka}, and {J. Luitz}}}]{Wien}
\bibinfo{author}{\bibnamefont{{P. Blaha}}}, \bibinfo{author}{\bibnamefont{{K.
  Schwarz}}}, \bibinfo{author}{\bibnamefont{{G.Madsen}}},
  \bibinfo{author}{\bibnamefont{{D.Kvasnicka}}}, \bibnamefont{and}
  \bibinfo{author}{\bibnamefont{{J. Luitz}}}, \emph{\bibinfo{title}{An
  augmented Plane Wave + Local Orbitals Program for Calculating Crystal
  Propertie}} (\bibinfo{publisher}{Karlheinz Schwarz, Techn. Universitat Wien,
  Austria, SBN 3-9501031-1-2.}, \bibinfo{year}{1999}).

\bibitem[{\citenamefont{{O. Gunnarsson} et~al.}(1989)\citenamefont{{O.
  Gunnarsson}, {O. K. Andersen}, {O. Jepsen}, and {J.
  Zaanen}}}]{Gunnarsson-hib}
\bibinfo{author}{\bibnamefont{{O. Gunnarsson}}},
  \bibinfo{author}{\bibnamefont{{O. K. Andersen}}},
  \bibinfo{author}{\bibnamefont{{O. Jepsen}}}, \bibnamefont{and}
  \bibinfo{author}{\bibnamefont{{J. Zaanen}}}, \bibinfo{journal}{Phys. Rev. B}
  \textbf{\bibinfo{volume}{39}}, \bibinfo{pages}{1708} (\bibinfo{year}{1989}).

\bibitem[{\citenamefont{{J. Allen, \emph{ et al.}}}(1981)}]{Allen}
\bibinfo{author}{\bibnamefont{{J. Allen, \emph{ et al.}}}},
  \bibinfo{journal}{Phys. Rev. Lett.} \textbf{\bibinfo{volume}{46}},
  \bibinfo{pages}{1100} (\bibinfo{year}{1981}).

\bibitem[{\citenamefont{{V. I. Anisimov} and {O.
  Gunnarsson}}(1991)}]{constrained-LDA}
\bibinfo{author}{\bibnamefont{{V. I. Anisimov}}} \bibnamefont{and}
  \bibinfo{author}{\bibnamefont{{O. Gunnarsson}}}, \bibinfo{journal}{Phys. Rev.
  B.} \textbf{\bibinfo{volume}{43}}, \bibinfo{pages}{7570}
  (\bibinfo{year}{1991}).

\end{thebibliography}

\end{document}